\documentclass[pra,twocolumn,showpacs,nobibnotes,aps]{revtex4-1}

\usepackage[utf8]{inputenc}
\usepackage{amsmath}
\usepackage{graphicx}
\usepackage{hyperref}

\newcommand{\braket}[2]{\langle #1 | #2 \rangle}

\newcommand{\dd}{\mathrm{d}}
\newcommand{\mdh}[1]{\frac{m #1}{\hbar^2}}
\newcommand{\A}[2]{{\mathcal{A}_{#1}^{[{#2}]}}}
\newcommand{\R}[1]{{\mathcal{R}_{#1}^{[k]}}}
\newcommand{\Ss}[2]{{\mathcal{S}_{#1}^{[{#2}]}}}
\newcommand{\rev}[1]{#1}

\newcommand{\fref}[1]{Fig.~\ref{#1}}

\newcommand{\eref}[1]{Eq.~(\ref{#1})}

\usepackage{color}

\definecolor{airforceblue}{rgb}{0.36, 0.54, 0.66}

\begin{document}

\preprint{APS/123-QED}

\title{Topological states in the Kronig--Penney model with arbitrary scattering potentials  }

\author{Irina Reshodko$^a$}
\email{irina.reshodko@oist.jp}

\author{Albert Benseny$^{a,b}$}

\author{\rev{Judit Romh\'{a}nyi$^{a}$}}

\author{Thomas Busch$^a$}

\affiliation{$^a$OIST Graduate University, 904-0495 Okinawa, Japan}
\affiliation{$^b$Department of Physics and Astronomy, Aarhus University, 8000 Aarhus C, Denmark}

\date{\today}

\begin{abstract}
We use an exact solution to the fundamental finite Kronig--Penney model with arbitrary positions and strengths of scattering sites to show that this iconic model can possess topologically non-trivial properties. 
By using free parameters of the system as extra dimensions we demonstrate the appearance of topologically protected edge states as well as the emergence of a Hofstadter butterfly-like quasimomentum spectrum, even in the case of small numbers of scattering sites.
We investigate the behaviour of the system in the weak and strong scattering regimes and observe drastically different shapes of the quasimomentum spectrum.

\end{abstract}


\maketitle

\section{\label{sec:intro}Introduction}

The Kronig--Penney (KP) model is one of the fundamental models of solid state physics and has since its inception~\cite{KronigPenney1931} received significant attention. It combines predictive power with accessibility and has, in fact, become a standard model that is taught in almost all solid state classes for undergraduate students. Despite its underlying simplicity that neglects interactions between the particles, it is particularly well suited to describe the behaviour of electrons in metals~\cite{KP_book,McKellarStephenson1987,ChoPrucnal1987,Exner1995,YuhWang1988,Holzer1988}.
More recently an experimental realization of the KP potential for ultracold atoms in optical lattices was proposed~\cite{LackiZoller2016} and demonstrated~\cite{WangRolston2018} .

One important aspect of the success of the KP model lies in its flexibility.
It allows to describe impurities or disorder in an easy and straightforward manner by assuming the scattering potentials to be located at non-periodic positions or having random strengths~\cite{Kerner1954,MihokovaSchulman2016}.
Here we present an analytical solution for the arbitrary finite KP model, when all scatterers are placed at arbitrary positions and have arbitrary strengths.
Due to the generality of the presented solution, the arbitrary finite KP model is broadly applicable for real systems, such as crystals made from multiple atomic species, impurities in the spacial periodicity with respect to position and scattering strength, and effects stemming from finite geometries. 
This can be used for the exact treatment of effects that were only explored numerically before, such as  localisation~\cite{Souillard1984,Delyon1984,Izrailev1999} or the existence of topologically non-trivial states~\cite{StJean2017,LangCaiChen2012}.
As an example we use the explicit solution in order to investigate appearance of the edge states and Hofstadter butterfly-like features in a finite, continuous system. 

To obtain the single-particle solutions of the arbitrary finite KP model we use the coordinate Bethe ansatz approach.
This method of solving one-dimensional quantum many-body problems was first described by Hans Bethe in 1931~\cite{Bethe1931}, and has since then been successfully applied to a large number of problems in lower dimensions~\cite{LiebLiniger1963, Lieb1963, Gaudin1971,Sutherland1968,Robinson2016}.

Our manuscript is organised as follows. In Sec.~\ref{derivations} we outline the solution to the arbitrary finite KP model using the Bethe ansatz. For this we first define the problem in Sec.~\ref{model} and then derive and present the explicit expressions for the Bethe equation and the eigenfunctions, in Secs.~\ref{derbeq} and  \ref{solution}, respectively. In Sec.~\ref{edgest} we use these solutions to demonstrate the appearance of edge states in the finite KP model with a lattice shift as an extra dimension. Finally, in Sec.~\ref{hofstadter}, we explore the emergence of a Hofstadter butterfly-like energy spectrum in the amplitude-modulated KP model.

\section{Solution of the arbitrary finite KP model}
\label{derivations}

\subsection{Model}
\label{model}

\begin{figure}[tb]
\includegraphics[width=\linewidth]{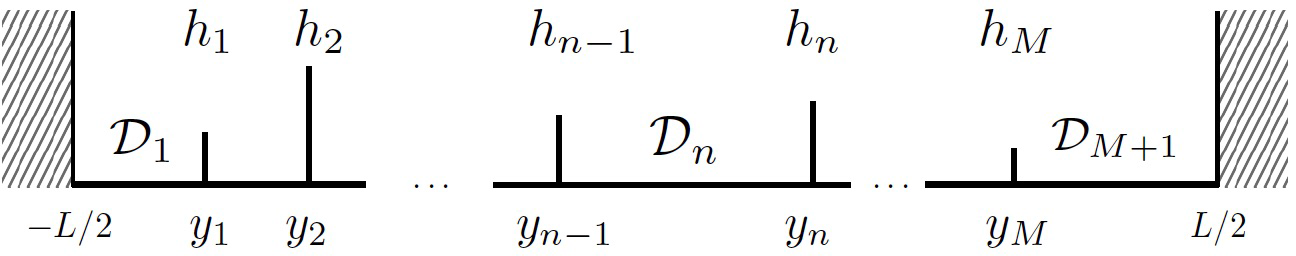}
\caption{Schematic of the arbitrary finite KP model. The barriers are located at positions $y_1$ to $y_M$ and have respective heights of $h_1,\dots,h_M$. The regions between all scatterers (including the walls) are denoted as ${\mathcal D}_n = (y_{n-1}, y_{n})$, where $y_0$ and $y_{M+1}$ are the left and the right wall.\label{fig:Schematic}
}
\end{figure}

We consider a one-dimensional system consisting of an infinite potential box of size $L$, in which $M$ point-like scatterers of arbitrary strengths $\vec{h}=(h_1,\ldots, h_M)$ are placed at arbitrary positions $\vec{y}=(y_1,\ldots,y_M)$ with $y_n\in[-\frac{L}{2}, \frac{L}{2}]$ and $y_n<y_m$ for $n<m$ (see the schematic in \fref{fig:Schematic}), i.e.,
\begin{equation}
\label{eq:Potential}
V(x) =
\begin{cases}
\sum\limits_{n=1}^{M}h_n \delta(x - y_n),&\textrm{for}~-\frac{L}{2} < x < \frac{L}{2},\\
\infty&\textrm{otherwise}.
 \end{cases}
\end{equation}
We show below that this potential cannot, in general, be solved with the Bethe ansatz for a system of point-like interacting bosons. However, the non-interacting and the infinitely strongly interacting (Tonks--Girardeau) limit can be solved, the latter by making use of the Bose--Fermi mapping theorem~\cite{Girardeau1960,Girardeau1965}.
For both it is necessary to consider only the single-particle Hamiltonian and the corresponding Schr\"{o}dinger equation
\begin{equation}
   \label{eq:scheq}
   -\frac{\hbar^2}{2m}\frac{\dd^2\Psi(x)}{\dd x^2} + V(x) \Psi(x) = E\Psi(x) .
\end{equation}

The essence of the coordinate Bethe ansatz approach is that the eigenstates of any system can be represented as a superposition of plane waves with different quasi-momenta for each particle~\cite{Korepin1993}.
By taking into account all possible scattering events one can construct a set of consistency equations, called the Bethe equations, and
only those quasi-momenta which satisfy the Bethe equations are allowed in the system.
Once the quasimomenta are determined, the energy of the system is given by a simple sum of their squares.
A necessary condition for a system to be integrable, however, is that it satisfies the Yang--Baxter relations~\cite{Sutherland2004,Jimbo1990}. 
These  stem from the requirement that all of the three-body scattering processes in the system can be decomposed into a series of two-body scattering events whose order does not matter.

Unfortunately, for larger numbers of particles and barriers, or for a non-symmetric placement of a single barrier, the Yang--Baxter relation cannot be satisfied in the regime of finite interactions.
This can be seen straightforwardly by considering two particles with different quasi-momenta located in the same region between two scatterers.
The three-body scattering events occur when both particles hit the same barrier at the same time. 
While these events can in principle be decomposed into the two particles scattering between themselves, and each particle individually scattering with the barrier, the order in which the particles scatter against the barrier matters. This is because the second particle to scatter on the barrier will be subject to a different dynamical evolution depending on the quasi-momentum of the other particle.
Consequently, the Yang--Baxter relations cannot be fulfilled and the model cannot be solved analytically with the Bethe ansatz for finite interaction strengths.

It is worth noting, however, that the interacting case was recently studied for a specific example by Liu and Zhang~\cite{LiuZhang2015}, who considered one scatterer at the centre of the infinite box ($M=1$, $y_1=0$).
They showed that this system can be partially solved for two particles and arbitrary scattering strengths, by finding the eigenstates for which the Yang--Baxter relations are satisfied.

Let us also mention that a different approach to the single-particle  problem was recently proposed by Sroczy\'nska {\it et al.}~\cite{Idziaszek2018}.
In this work the authors use a Green's function approach to solve the problem of a single-particle moving in an arbitrary trapping potential which has regularized delta scatterers superposed.
The solution presented here will coincide with the one-dimensional solution of Ref.~\cite{Idziaszek2018} after substitution of the Green's function for the infinite square well.

\subsection{Bethe equations}
\label{derbeq}
In the following we outline the exact solution of eq.~\eqref{eq:scheq}, with the potential given by expression \eqref{eq:Potential}, using the Bethe ansatz. 
We start by considering solutions for each of the regions between the scatterers, determined by the free-particle Schr\"odinger equation
\begin{equation}
\label{eq:freepart}
-\frac{\hbar^2}{2m}\frac{\dd^2\Psi(x)}{\dd x^2}=E \Psi(x)
\end{equation}
We construct an ansatz for the full solution of eq.~\eqref{eq:freepart} composed of piecewise plane waves with quasi-momentum $k$ as
\begin{equation}
\label{eq:ansatz}
\Psi(x) =\sum\limits_{n=1}^{M+1}\left( \A{n}{k}e^{i k x}+\A{n}{-k}e^{-i k x}\right)\Theta_n(x),
\end{equation}
where $\Theta_n(x) \equiv \theta(x-y_{n-1})\theta(y_n-x)$, with $\theta(x)$ being the Heaviside step function. Each term in the sum corresponds to a region between two scatterers or a scatterer and the adjacent walls, $x \in {\mathcal D}_n = (y_{n-1}, y_{n})$, and we have set $y_0= -L/2$ and $y_{M+1}=L/2$.
This ansatz has to satisfy the boundary conditions imposed by the scatterers and walls, which are of the form 
\begin{align}
\label{eq:bcboxedge}
\Psi(\pm L/2) &= 0,\\
\label{eq:barrcont}
\left.\Psi(x)\right|_{x\to y_n^-}&=\left.\Psi\right|_{x\to y_n^+} , 
\\
\label{eq:barrscat}
\frac{2 m h_n}{\hbar^2}\Psi(y_n) &= \left.\frac{\dd\Psi(x)}{\dd x}\right|_{x\to y_n^+}-\left.\frac{\dd\Psi(x)}{\dd x}\right|_{x\to y_n^-} .
\end{align}
The Bethe ansatz approach now consists of constructing equations for the quasi-momenta $k$, from which the energies follow as $E = \hbar^2 k^2/2 m$.
To do so, we will first construct expressions for all coefficients $\A{n}{k}$ and $\A{n}{-k}$ by calculating for each region $\mathcal{D}_n$ the elements of the reflection matrix
\begin{equation}
\R{n} \equiv \frac{\A{n}{-k}}{\A{n}{k}}.
\label{eq:refDef}
\end{equation}

Starting with the boundary conditions at the walls, we first substitute the ansatz in eq.~\eqref{eq:ansatz} into eq.~\eqref{eq:bcboxedge},  and obtain expressions for the first and the last elements of the reflection matrix of the form
\begin{align}
\label{eq:wallleft}
\R{1}&=-e^{-i k L},\\
\R{M+1}&=-e^{i k L}.
\label{eq:wallright}
\end{align}
Next, the continuity and scattering conditions given in eqs.~\eqref{eq:barrcont} and \eqref{eq:barrscat} at the $j$-th barrier lead to
\begin{align}
\label{eq:bcontrel}
&\A{j+1}{k} + \A{j+1}{-k}e^{-i 2 y_j k} = \A{j}{k} + \A{j}{-k}e^{-i 2 y_j k} ,
\\
\label{eq:bscatrel}
&\A{j+1}{k}\left(i k - \mdh{h_j}\right) - \A{j+1}{-k}e^{-i 2 y_j k}\left(i k + \mdh{h_j}\right)  \nonumber \\
&\quad=\A{j}{k}\left(i k + \mdh{h_j}\right) + \A{j}{-k}e^{-i 2 y_j k}\left(-i k + \mdh{h_j}\right).
\end{align}
By considering eqs.~\eqref{eq:bcontrel} and \eqref{eq:bscatrel} for $j=n, n-1$ and taking into account that $\A{j}{-k}=\R{j}\A{j}{k}$, we find the recursive form for the reflection matrix elements 
\begin{align}
\label{eq:reflectleft}
\R{n} &=\frac{-\mdh{h_{n-1}}e^{i 2 y_{n-1} k} + [i k -\mdh{h_{n-1}}]\R{n-1}}{[i k + \mdh{h_{n-1}}] + \mdh{h_{n-1}}e^{-i 2 y_{n-1} k}\R{n-1}},
\\
\label{eq:reflectright}
\R{n} &= \frac{\mdh{h_n}e^{i 2 y_n k}+ [i k + \mdh{h_n}]\R{n+1}}{[i k - \mdh{h_n}] - \mdh{h_n} e^{-i 2 y_n k}\R{n+1}},
\end{align}
for the inner regions $\mathcal{D}_n, n=2\dots M$.
Together with eqs.~\eqref{eq:wallleft} and \eqref{eq:wallright},  these expressions correspond to the two ways of inverting the sign of the quasi-momentum $k$ by reflecting the particle at the left or the right wall.
Thus, the two expressions for each region have to be equivalent, yielding  the $M+1$ Bethe equations that define the allowed quasi-momenta of the system.

Next we prove that all these Bethe equations are equivalent.
For this we represent the process of reflecting a particle as a sequence of scattering events at each barrier, denoted by the elements of the scattering matrix
\begin{equation}
\Ss{n}{\pm k} \equiv \frac{\A{n+1}{\pm k}}{\A{n}{\pm k}} ,
\end{equation}
and reflections against the left and right wall, denoted by $\R{1}$ and $\R{M+1}$.
For example, reflecting the particle from the rightmost region, $\mathcal{D}_{M+1}$, the Bethe equation can be written as
\begin{equation}
\prod\limits_{n=1}^{M}\Ss{n}{-k}^{-1} \times \R{1} \times \prod\limits_{s=1}^{M}\Ss{M+1-s}{k} = \R{M+1}.
\end{equation}
It is easy to see that by multiplying both sides of the equation by the inverse scattering matrices in the appropriate sequence, one can reconstruct similar equations for all other regions.

Consequently, we only need one Bethe equation for the single variable $k$, and we choose the one that assumes the particle to be in the rightmost region, as it has the simplest form given by
\begin{equation}
\label{eq:BE}
\frac{-\mdh{h_{M}}e^{i 2 y_{M} k} + [i k -\mdh{h_{M}}]\R{M}}{[i k + \mdh{h_{M}}] + \mdh{h_{M}}e^{-i 2 y_{M} k}\R{M}}=-e^{i k L}.
\end{equation}
By unwrapping the this recursive expression, we can then construct the Bethe equation for any given values of the system parameters, which can be algebraically simplified to
\begin{equation}
0 = \sum_{n=0}^M \left(\frac{2m}{\hbar^2}\right)^{n} \xi_n k^{M-n},
\end{equation}
where
\begin{equation}
\xi_n = \sum_{\substack{(p_1< \ldots< p_n) \\ 1\le p_i \le M}}
 \left( \prod_{j=1}^{n} h_{p_j} \right)
 \left( \prod_{j=1}^{n+1} \sin \left[k(y_{p_j} - y_{p_{j-1}})\right] \right) . 
\end{equation}
The sum is over all ordered sets of $n$ scatterer indices, i.e., $(p_1, \ldots, p_n)$, and we have also defined $y_{p_0} = -L/2$ and $y_{p_{n+1}}=L/2$.
The Bethe equation constructed in this way is an algebraic transcendental equation, whose roots can generally only be found via numerical methods or analytical methods for small number of roots~\cite{Luck2015}.

\subsection{The wavefunction}
\label{solution}
From eqs.~\eqref{eq:bcontrel} and \eqref{eq:bscatrel} we can also obtain an explicit recursive expression for the elements of the scattering matrix of the form
\begin{equation}
\Ss{n}{k} \equiv \frac{\A{n+1}{k}}{\A{n}{k}}
=1 + \frac{i}{k} \mdh{h_n} \left(1 +  e^{-i 2 y_{n} k}\R{n}\right),
\end{equation}
for $n=1\dots M$.
A similar expression can be obtained for $\Ss{n}{-k}$, however there is no need to  calculate it explicitly, 
as it can always be reconstructed from $\Ss{n}{k}$ and the reflection matrix.

We therefore have everything to express all coefficients of the ansatz wavefunction in eq.~\eqref{eq:ansatz} in terms of $\A{1}{k}$ as
\begin{align}
\A{n}{-k} &= \R{n} \times \A{n}{k}, \\
\A{n}{k} &= \prod\limits_{j=1}^{n-1}\Ss{j}{k} \times \A{1}{k}.
\end{align}
The remaining coefficient $\A{1}{k}$ is in principle fixed by normalization of the wavefunction.
The explicit form of the coefficients is then given by
\begin{align}
\A{n}{k} &= e^{i k \frac{L}{2} }\left(1 + \sum\limits_{j=1}^{n-1}\left(\frac{2 m}{k \hbar^2}\right)^j\Xi_j^{n}(k)\right)\A{1}{k}, \\
\A{n}{-k} &= - \A{n}{k}^*,
\end{align}
where
\begin{equation}
\Xi_j^n=\sum_{\substack{(p_1, \ldots, p_j) \\ 1\le p_i \le n-1}} e^{- i k (y_{p_j} +\frac{L}{2})} \prod\limits_{l=1}^j h_{p_l}\sin{[k (y_{p_{l}}-y_{p_{l-1}})]}.
\end{equation}
For practical purposes we define $\A{1}{k}=1$ and later renormalize all coefficients $\A{n}{k}\to \mathcal{N} \A{n}{k}$
with the normalization constant
\begin{align}
\mathcal{N} &=
\Bigg[\sum\limits_{n=1}^{M+1} \Bigg(
2|\A{n}{k}|^2(y_n-y_{n-1})
\\ \nonumber & \qquad
- \frac{\sin 2 k y_n - \sin 2 k y_{n-1}}{2k} (\Re \A{n}{k}^2 - \Im \A{n}{k}^2)
\\ \nonumber & \qquad
- \frac{\cos 2 k y_n - \cos 2 k y_{n-1}}{k}\Re \A{n}{k} \Im \A{n}{k}
\Bigg) \Bigg]^{-\frac{1}{2}},
\end{align}
where $\Re$ and $\Im$ denote the real and imaginary parts.

\section{Nontrivial topology and boundary states}
\label{edgest}
Since now we have access to all single-particle states of the system, we will apply our solution for the cases equidistant scatterers of equal and varying heights. 
Our results are detailed in Sec. \ref{sec:uniform} and \ref{sec:nonuniform}, respectively.

For both cases we study the emergent topology of the bands. 
Even though topological effects generally require higher dimensions, it has been shown that in certain one-dimensional systems  additional degrees of freedom  can be used as a virtual second dimension - a \textit{superspace}~\cite{Lange1984}.
The Fermi-Hubbard model with modulated on-site energy~\cite{LangCaiChen2012},  and the non-interacting system with trigonometric potential~\cite{ZhengYang2014} provide examples for nontrivial topology studied in 1+1 dimensions.
\rev{A nontrivial topology is marked by a nonzero topological invariant, such as the first Chern number in the case of two dimensional systems~~\cite{Resta1994, Xiao2010}. }

To have a second periodic parameter that can be considered as a virtual dimension, we apply a shift $\Delta$ to the scatterers with respect to the wall of the box.

\begin{figure}[tb]
\includegraphics[width=\linewidth]{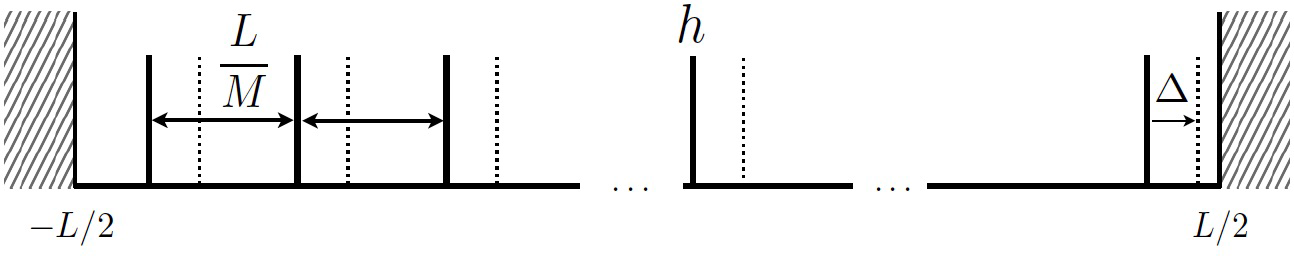}
\caption{
Schematic of the arbitrary finite KP model with equidistant barriers of equal height $h$.
The superspace is realised by introducing a lattice shift $\Delta$, which displaces the barriers to new positions (dotted lines), 
\label{fig:schemShift}
}
\end{figure}

First we use our analytical solution for the finite models.  
Plotting the energies with respect to the shift $\Delta$, we find in-gap modes connecting the bands. 
We then investigate the appearance of edge states for two paradigmatic structures: a uniform lattice and a superlattice.  
In both cases the superspace is realized by a relative shift of the lattice of scatterers with respect to the box (see \fref{fig:schemShift}).

\subsection{Equidistant scatterers of equal heights}
\label{sec:uniform}

\begin{figure}
\begin{minipage}{0.4\linewidth}
{(a)~spectrum}
\includegraphics[width=0.99\linewidth]{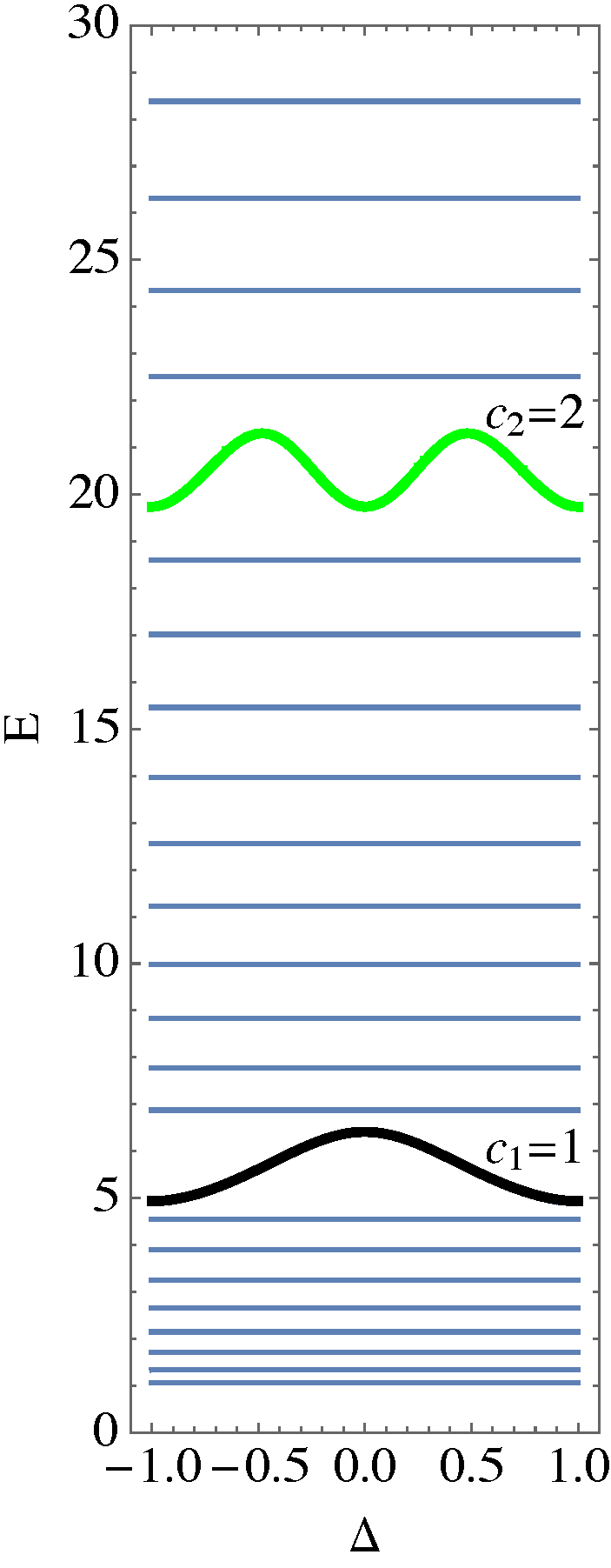}
\end{minipage}
\   \hfill
\begin{minipage}{0.5\linewidth}
{(b)~second edge state  (green)}
\includegraphics[width=0.99\linewidth]{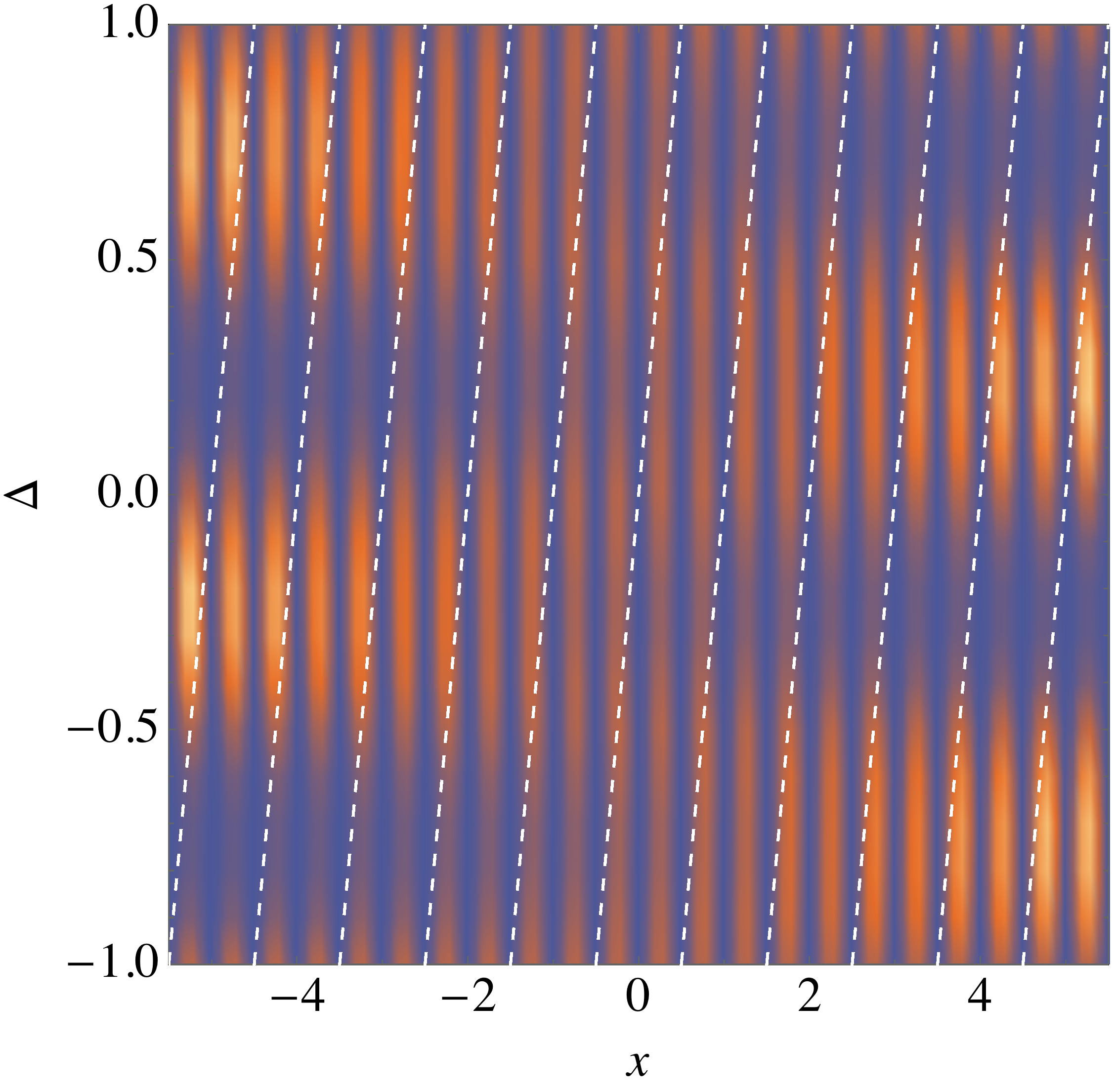}
{(c)~first edge state (black)}
\includegraphics[width=0.99\linewidth]{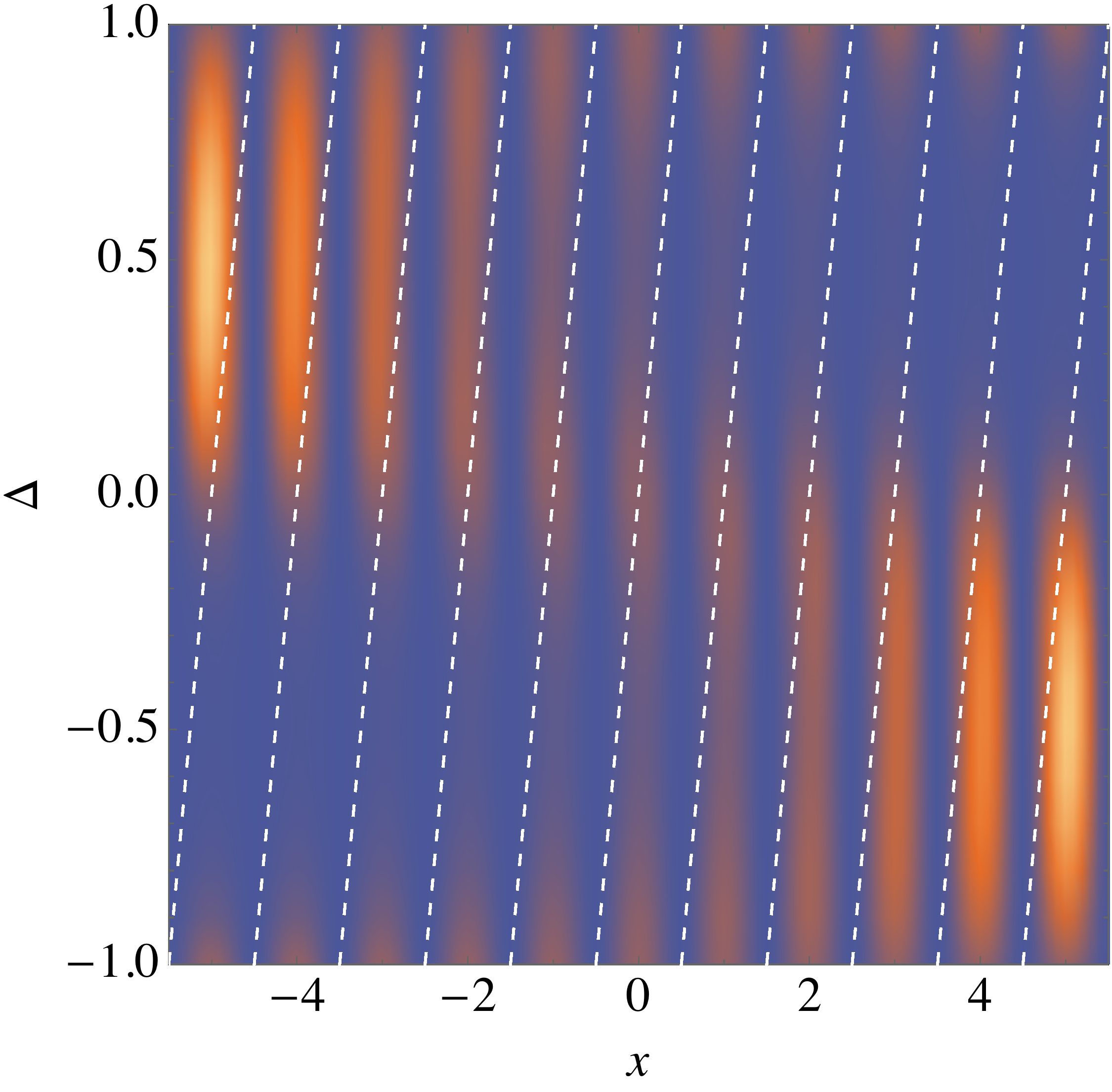}
\end{minipage}
\caption{
\label{fig:sameheights}
(a) \rev{Energy spectrum as a function of the shift $\Delta$ for a system of 11 equidistant barriers of height $h=0.4$ in a box of size $L=11$.
The numbers in the band gaps $c_1$ and $c_2$ are the Chern numbers of the underlying band.}
The green and black lines indicate the first two edge states with quantum numbers 22 and 11, whose densities are shown in (b) and (c), respectively.
The dashed white lines indicate the positions of the scatterers.
}
\end{figure}

We first consider a set of equidistant scatterers of equal height $h_n = h>0$ for $n=1,..,M$ and introduce a shift in the barrier positions $\Delta \in [-1,1]$ with respect to the walls of the  box
\begin{equation}
\label{eq:edgest_pos}
y_n = - \frac{L}{2} + \left(n  +\frac{\Delta-1}{2} \right)\frac{L}{M}. 
\end{equation}
To simplify the presentation of our results, we will use natural units, $m=\hbar=1$, from now on.
The energy spectrum as a function of $\Delta$ is shown in \fref{fig:sameheights}(a), where one can clearly see the appearance of in-gap states between the bands, even in the case of a rather small number of barriers.
We also show the probability density of the first two edge states as a function of $\Delta$, demonstrating localization of the two wavefunctions (\fref{fig:sameheights}(b,c)).
\rev{The strongest localization of the wavefunction is achieved for $k$ values in the middle of the band gap ($\Delta = \pm \frac{1}{2}$), showing that the in-gap states indeed live on the edges.  
The density becomes stretched over the whole box for the shift values when the in-gap states approach the bulk bands, e.g. for $\Delta \rightarrow 0$. This indicates that here the edge modes submerge in the bulk. 
The slope of the energy of the edge states as a function of the shift $\Delta$ denotes their velocities: the edge modes with positive slope have positive velocity, and the ones with negative slope have negative velocity. 
From \fref{fig:sameheights} (b) and (c) it is clear that the edge modes with opposite velocities, i.e. traveling in opposite directions, are located on the opposite sides of the system, signaling that indeed these boundary states are chiral. 
In the second gap we have two edge modes on both sides. 
From \fref{fig:sameheights} it is evident that those on the same side propagate in the same direction, and therefore cannot cancel each other. 
In-gap chiral edge modes suggests the presence of nontrivial topology. 
In order to prove this, we numerically calculate the Chern number of the first two energy bands in the system with periodic boundary conditions using the method described in~\cite{Fukui2005}.
The Chern number is a topological invariant connected to the Berry phase and defined as $c = \frac{1}{2 \pi} \int \limits_{k} \textrm{d} k\int \limits_{\delta} \textrm{d} \delta \left(\partial_{k}A_{\delta} - \partial_{\delta}A_{k}\right)$, where $k$ is the quasi-momentum in $x$ direction and $\delta$ is the lattice shift. 
$A_{k_1} = i \braket{\phi(k_1, k_2)}{\partial_{k_1} \phi(k_1, k_2)}$ and $A_{k_2}=i \braket{\phi(k_1, k_2)}{\partial_{k_2} \phi(k_1, k_2)}$ are the Berry connections with $\phi(k_1, k_2)$ being the occupied Bloch state~\cite{Resta1994, Xiao2010}.
A topologically nontrivial system will have a non-zero integer Chern number.
Our numerical calculation of the Chern numbers for the first two energy bands are shown in \fref{fig:sameheights}, and are both equal to one.
According to the bulk-boundary correspondence, the number of edge modes in the gap has to be related to the sum of the Chern numbers of the bands up to the given gap. 
In the first gap we have one edge mode with positive and one with negative velocity, located on the opposite ends and corresponding to $c_1=1$. 
In the second gap the number of edge modes becomes two for both sides, reflecting a total  Chern number $c_2=2$, and so on. }
\subsection{Equidistant scatterers of non-uniform heights}
\label{sec:nonuniform}
\begin{figure}
\begin{minipage}{0.4\linewidth}
{(a)~spectrum}
\includegraphics[width=0.99\linewidth]{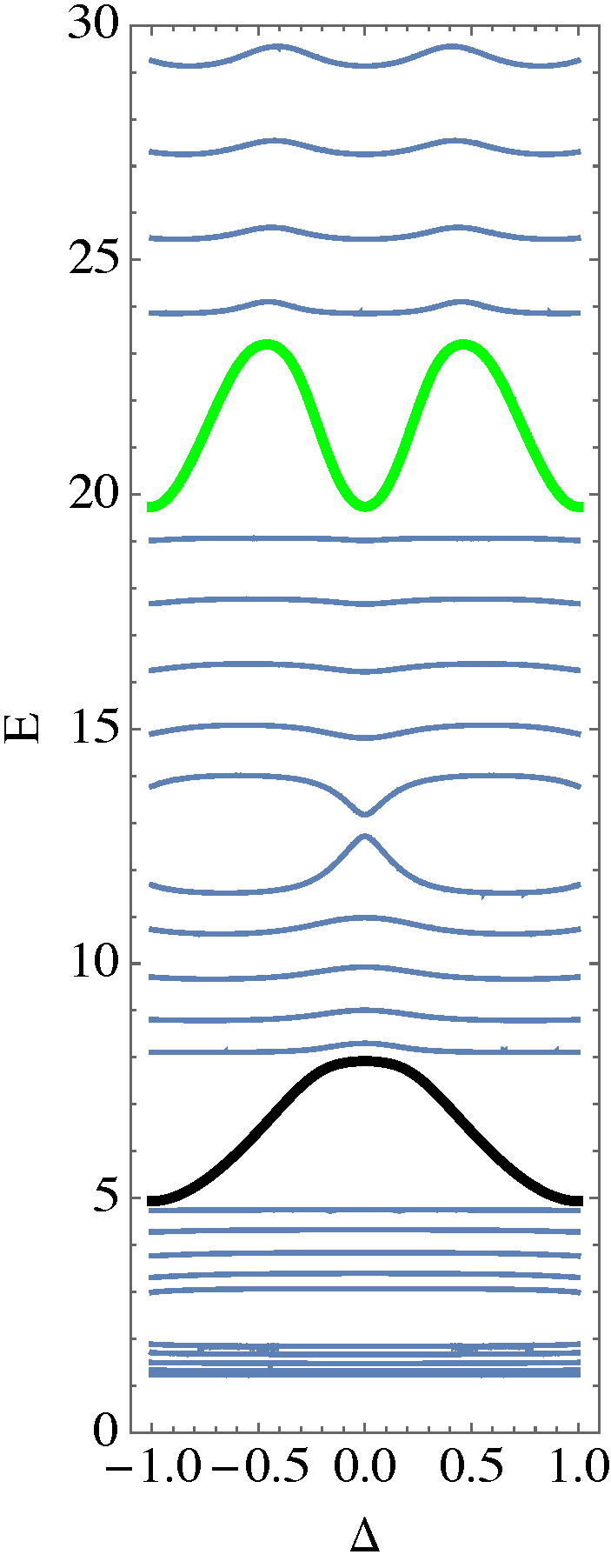}
\end{minipage}
\   \hfill
\begin{minipage}{0.5\linewidth}
{(b)~second edge state (green)}
\includegraphics[width=0.99\linewidth]{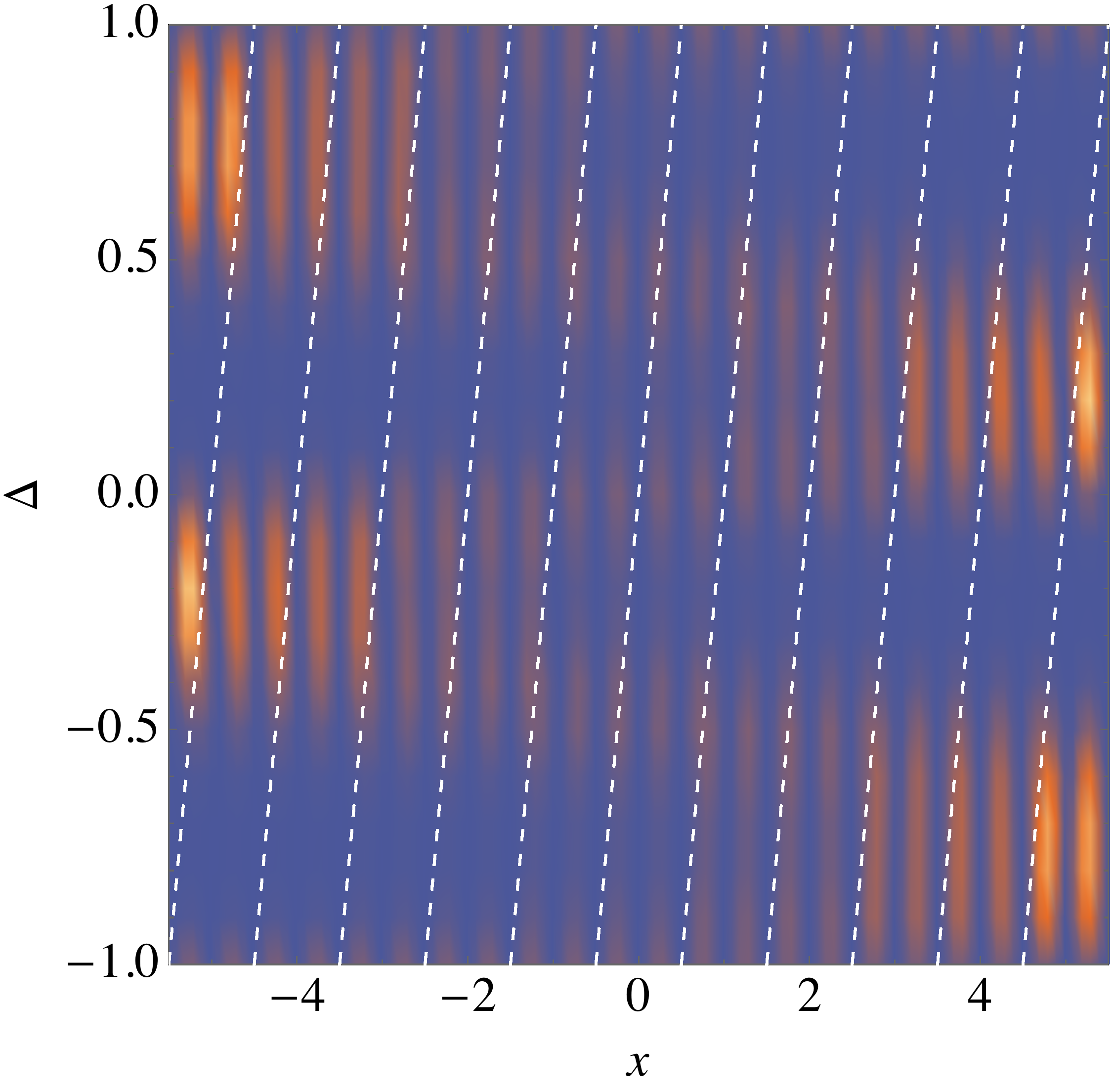}
{(c)~first edge state (black)}
\includegraphics[width=0.99\linewidth]{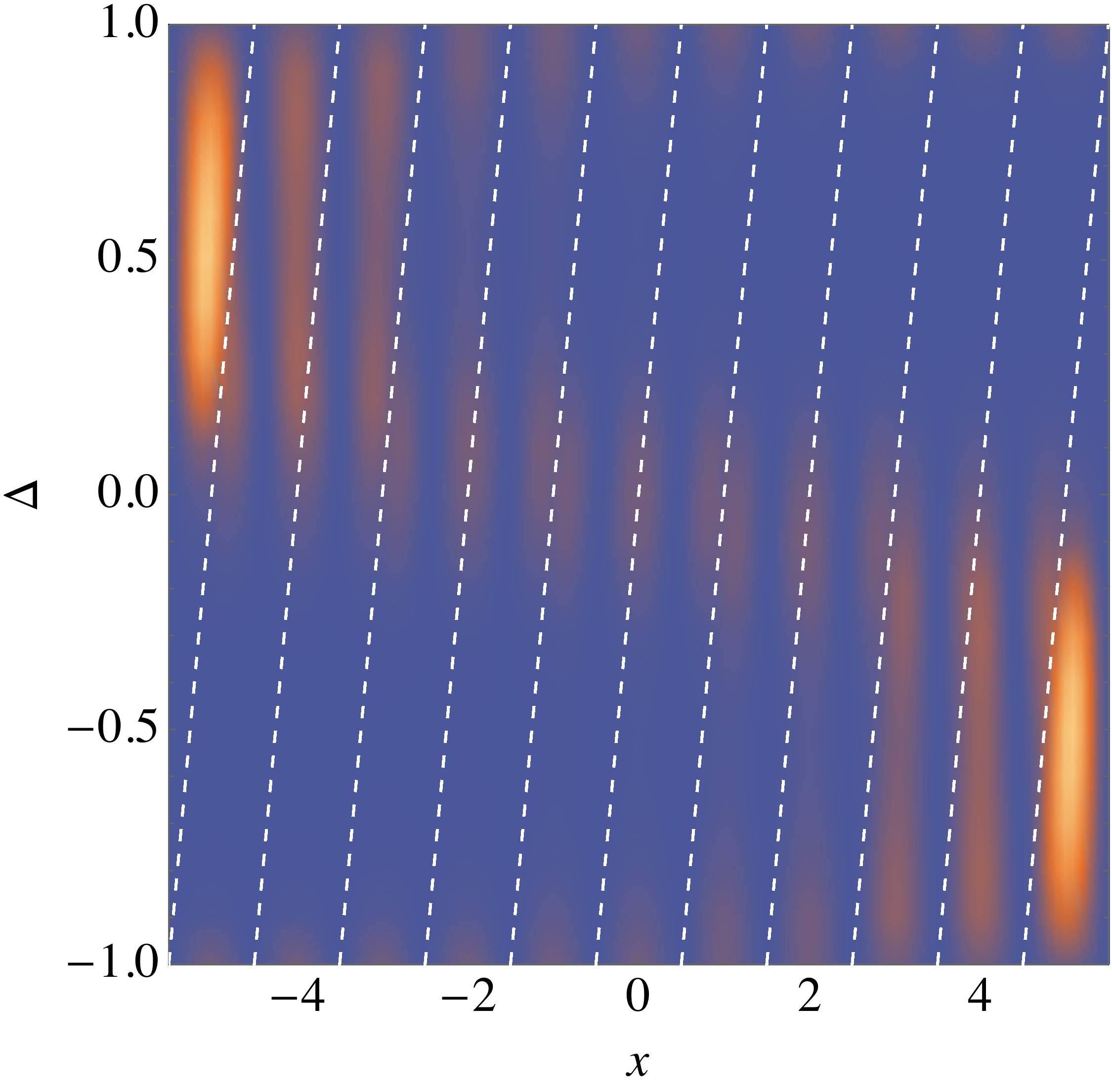}
\end{minipage}
\caption{
\label{fig:alterheights}
Same as \fref{fig:sameheights}, but for a system of 11 equidistant barriers of  alternating heights $h=\{0.4, 1.4\}$.}
\end{figure}

\begin{figure}
\begin{minipage}{0.4\linewidth}
{(a)~spectrum}
\includegraphics[width=0.99\linewidth]{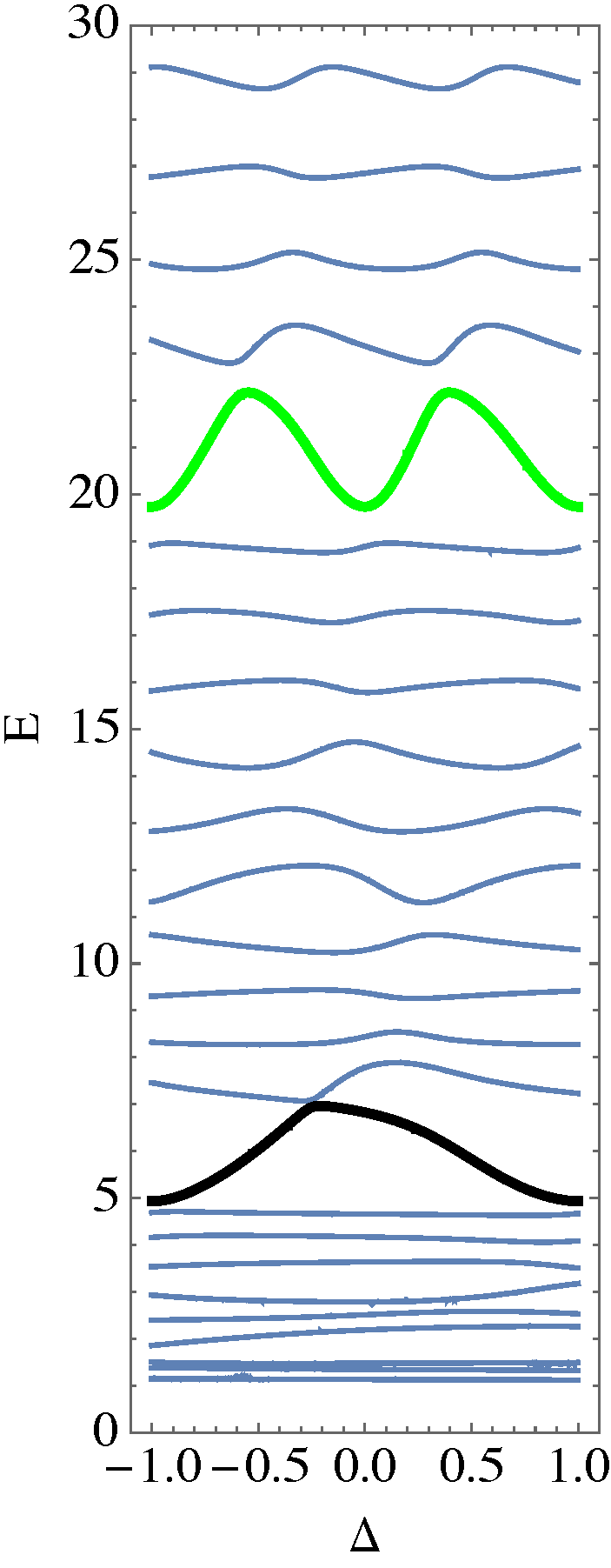}
\end{minipage}
\   \hfill
\begin{minipage}{0.5\linewidth}
{(b)~second edge state  (green)}
\includegraphics[width=0.99\linewidth]{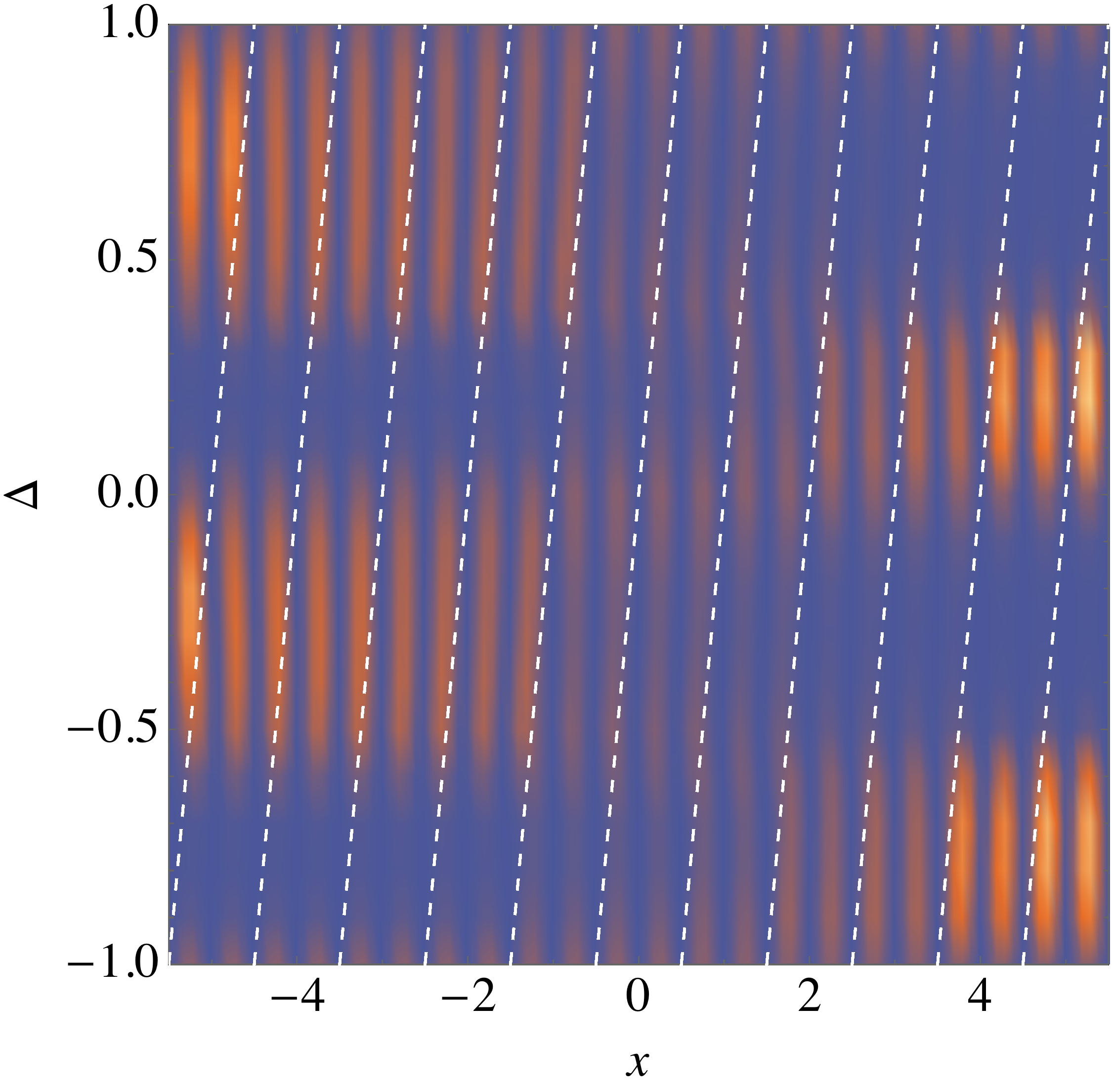}
{(c)~first edge state (black)}
\includegraphics[width=0.99\linewidth]{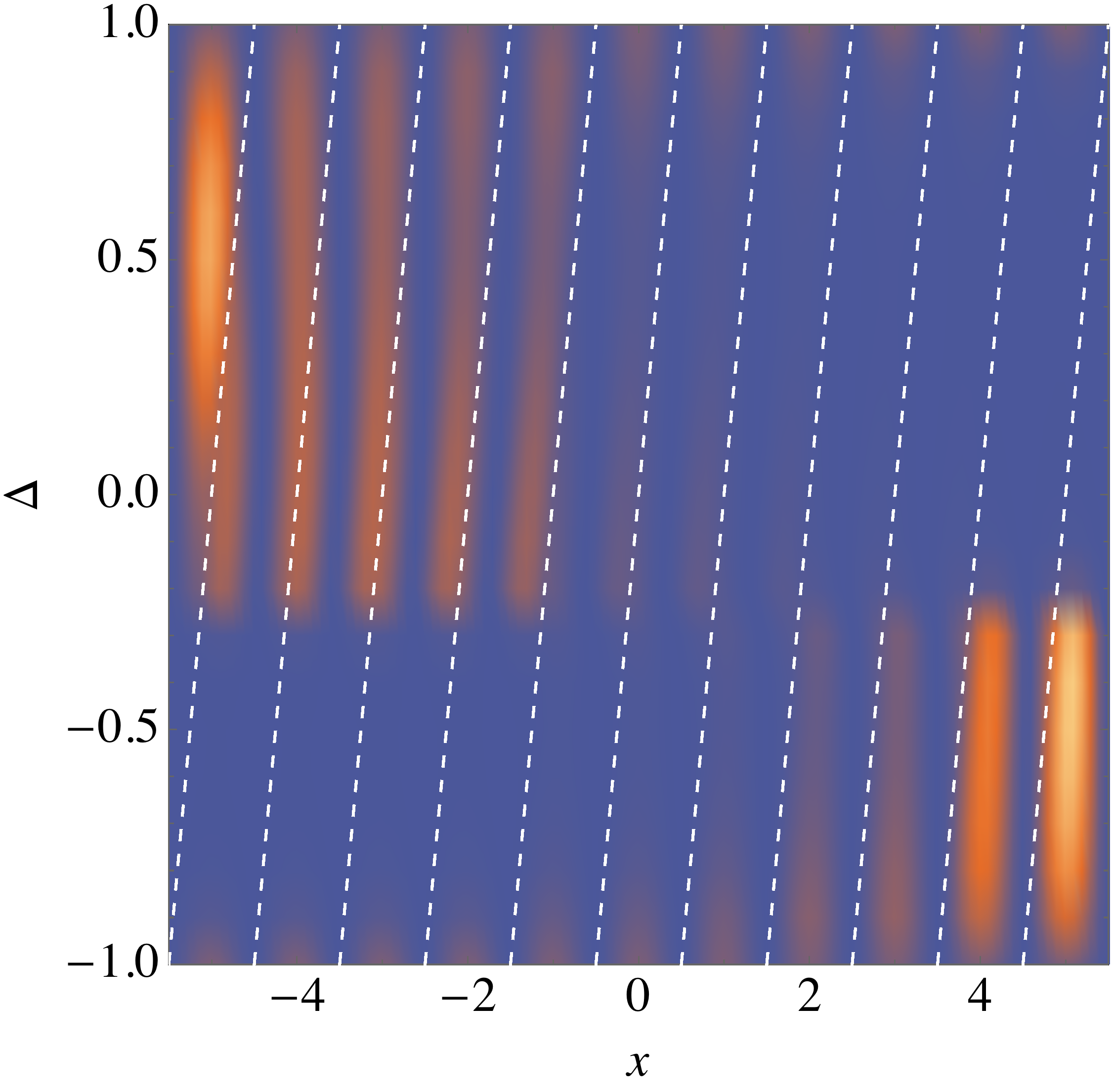}
\end{minipage}
\caption{
\label{fig:randheights}
Same as \fref{fig:sameheights}, but for an example of a system of 11 equidistant barriers of random heights varying from  $h_\textrm{min}=0.1$ to $h_\textrm{max}=1.4$.
}
\end{figure}

While superlattice-type structures are more complicated, they can still be treated straightforwardly using the above solution. 
Here we consider again a system of equidistant scatterers (cf.~\eref{eq:edgest_pos}), but with alternating heights $h=\{0.4, 1.4\}$. 
As expected, the energy spectrum becomes more complicated with additional gaps appearing (see \fref{fig:alterheights}), which are due to the existence of two different sub-lattices~\cite{Eldib1987}.
\rev{Adiabatically altering the heights of the potentials does not close the gaps observed for the uniform case, therefore the topology of the bands cannot change. 
The chiral boundary states and the topological properties are robust against perturbing the system with potentials of random heights, as shown in  \fref{fig:randheights}.}

\section{Hofstadter butterfly and cocoon spectra}
\label{hofstadter}

Another characteristic topological effect is the appearance of a fractal pattern in the energy spectrum of a system  \cite{Hofstadter1976}.
This was first predicted by Hofstadter for electrons on an infinite 2D lattice in the presence of a magnetic field, where the particles experience a phase shift $\phi$ due to the magnetic field after a full loop over a lattice plaquette.
Such an energy spectrum has since then become known as a Hofstadter butterfly due to its distinct shape.
For finite systems, however, the fractal nature of the energy spectrum is known to be lost \cite{Blundell2004, Agarwala2017, Czajka2006}, but the overall shape of the butterfly is preserved, with states appearing in the bandgaps.
In one-dimensional systems similar effects can be observed when using a superspace\cite{Blundell1994, LangCaiChen2012}, and
here we will investigate the Hofstadter butterfly-like quasi-momentum spectrum of the arbitrary finite KP model as it emerges with increasing numbers of scatterers.

The model we are considering consists of equidistant barriers at positions $y_n = -L/2 + anL$, with $a = 1/{(M+1)}$.
The heights of the scatterers are modulated by a periodic function
\begin{equation}
h_n = h_\textrm{min} + (h_\textrm{max}-h_\textrm{min})\cos^2\left(2 \pi \phi\left(a n + \frac{1}{2}\right)\right). 
\label{eq:cosine}
\end{equation}

The scattering potential is periodic in $\phi$ (with period $\phi_0 \equiv (M+1)/2$), and $\phi$ therefore plays the role of the flux from the original Hofstadter study. 
Note that our model describes a continuous system, whereas the original Hofstadter argument was made for a system in the tight-binding approximation \cite{Hofstadter1976}.

\begin{figure}[tb]
  \begin{minipage}{0.48\linewidth}
  {(a)}
  \includegraphics[width=0.99\linewidth]{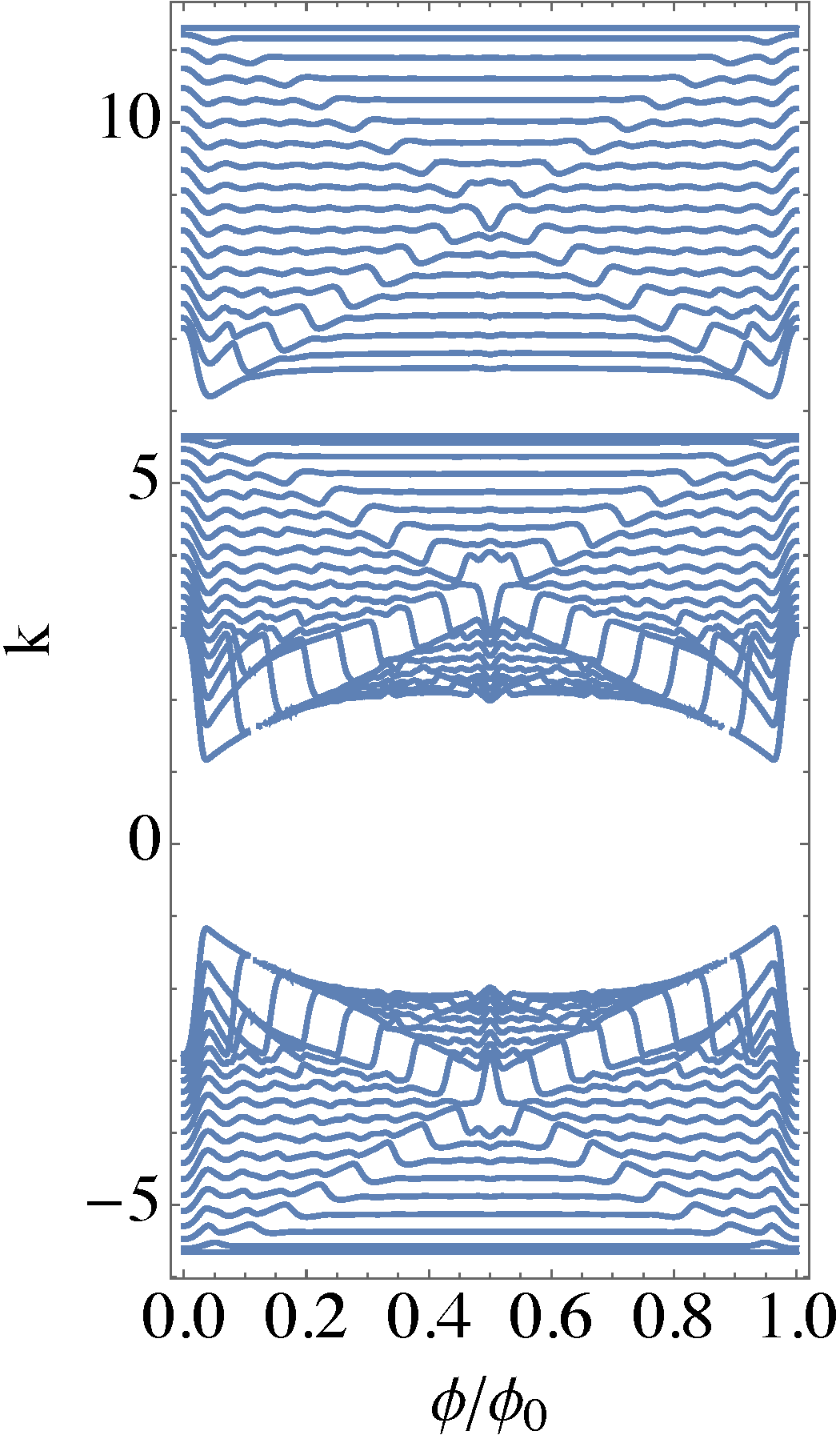}
   \end{minipage}   
  \begin{minipage}{0.48\linewidth}
  {(b)}
  \includegraphics[width=0.99\linewidth]{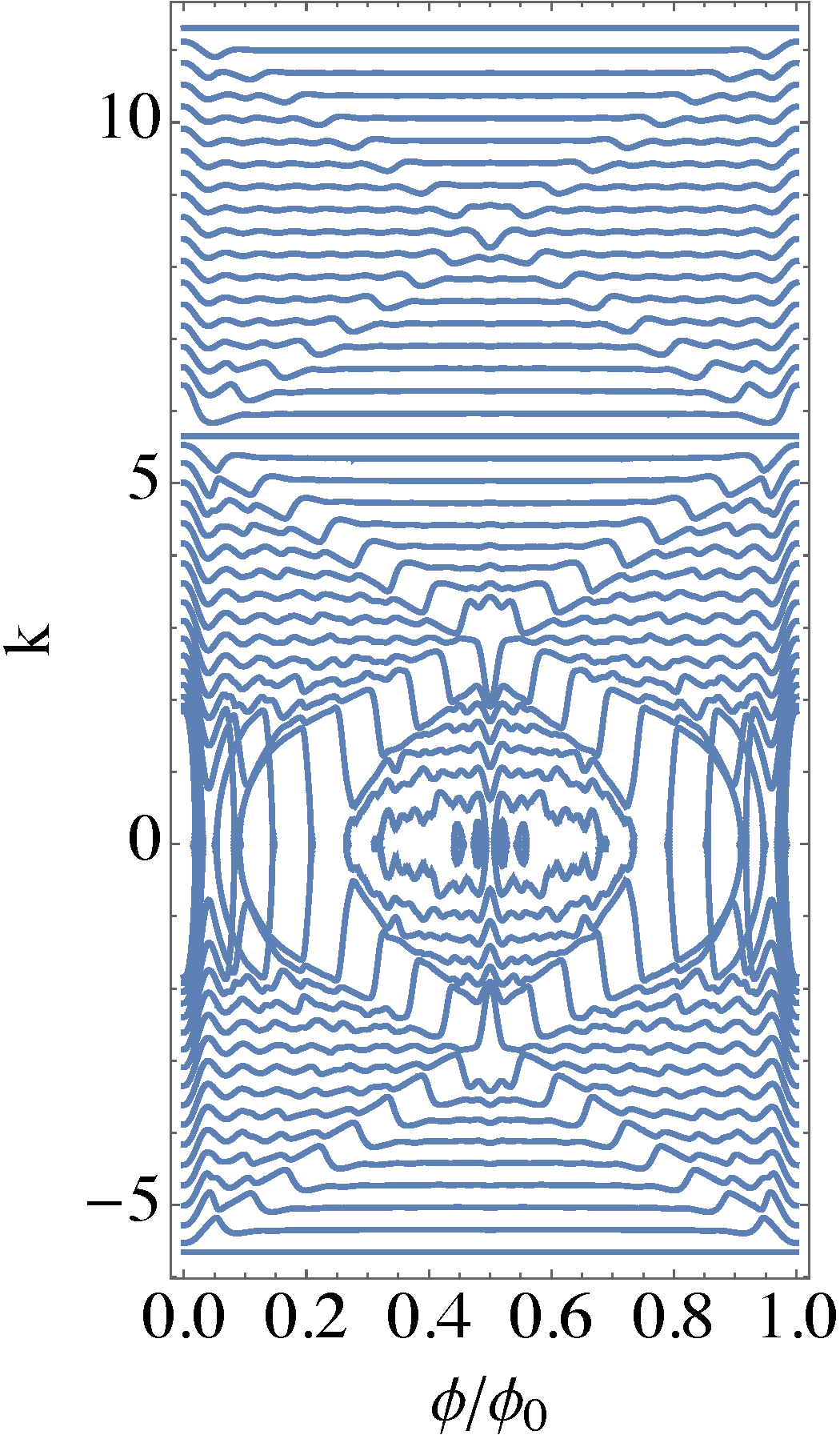}
  \end{minipage}
\caption{
  \label{fig:manyflies}
  (a) Hofstadter butterfly-like momentum spectrum in a system where all scatterer heights are positive.
  The modulation period is given by $\phi_0=\frac{M+1}{2}$, and the seventeen scatterer heights in this example vary between $h_\textrm{min} = 0.1$ and $h_\textrm{max} = 1.5$.
(b)  Same as above, but for a system with scatterer heights varying between $h_\textrm{min} = -0.5$ and  $h_\textrm{max} = 0.5$.
 The large circular feature in the center (a {\it cocoon}) is not present in the positive scatterers-only system, while the iconic {\it wings} are just developing.
}
\end{figure}

\begin{figure}[tb]
\begin{minipage}{0.48\linewidth}

{(a)~$M=5$}
\includegraphics[width=0.99\linewidth]{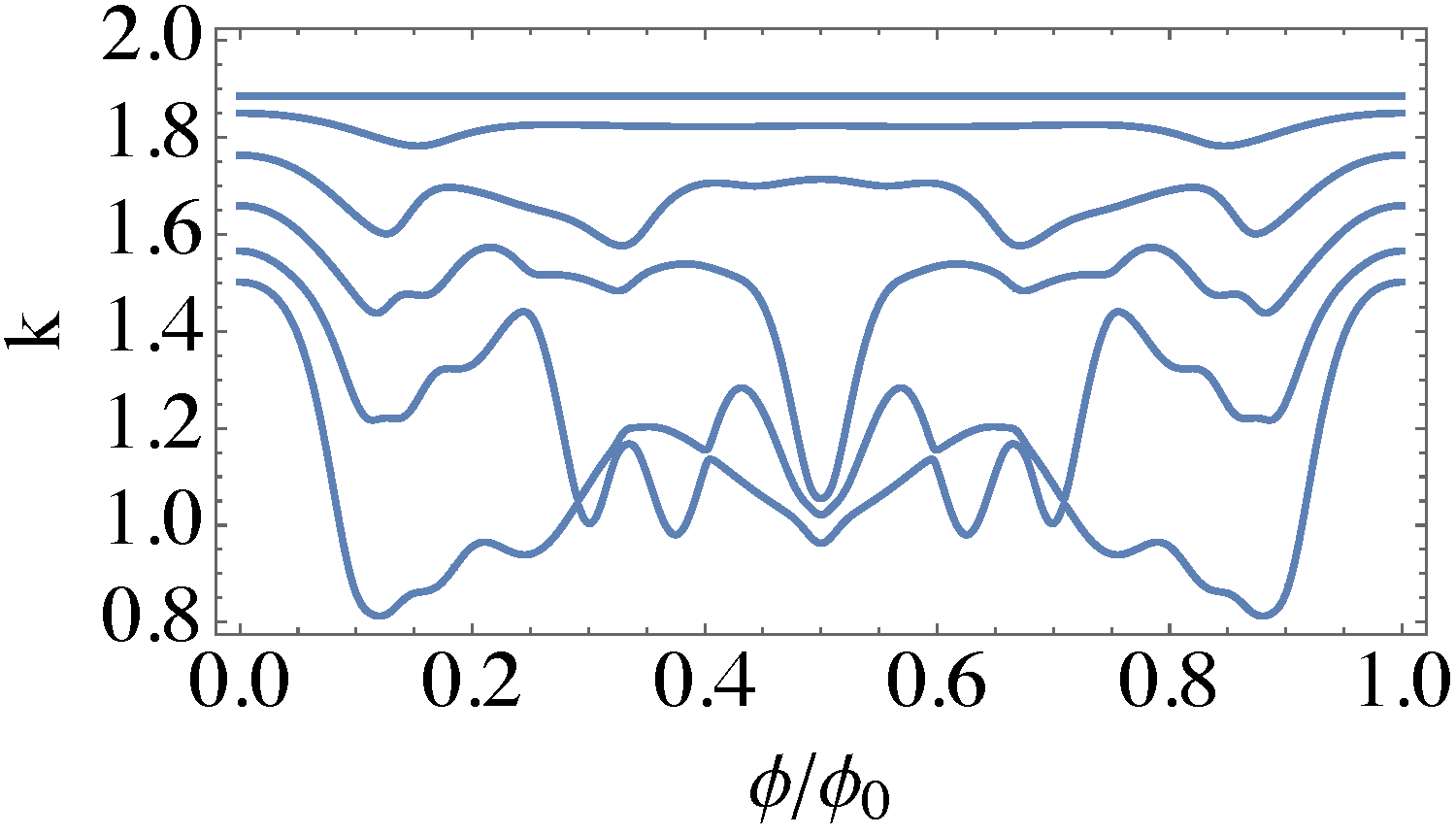}
{(c)~$M=11$}
\includegraphics[width=0.99\linewidth]{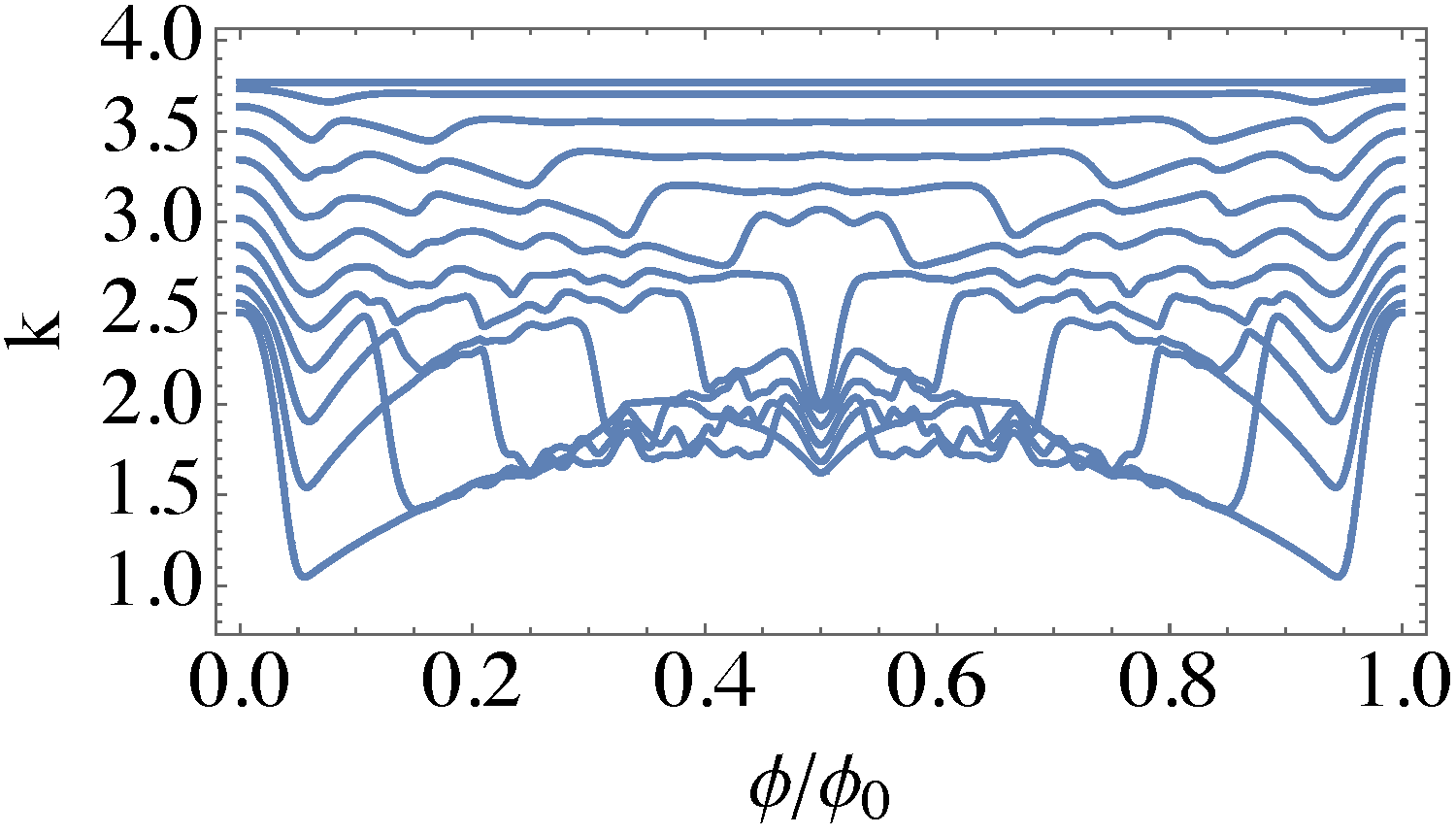}
\end{minipage}
\   \hfill
\begin{minipage}{0.48\linewidth}
{(b)~$M=7$}
\includegraphics[width=0.99\linewidth]{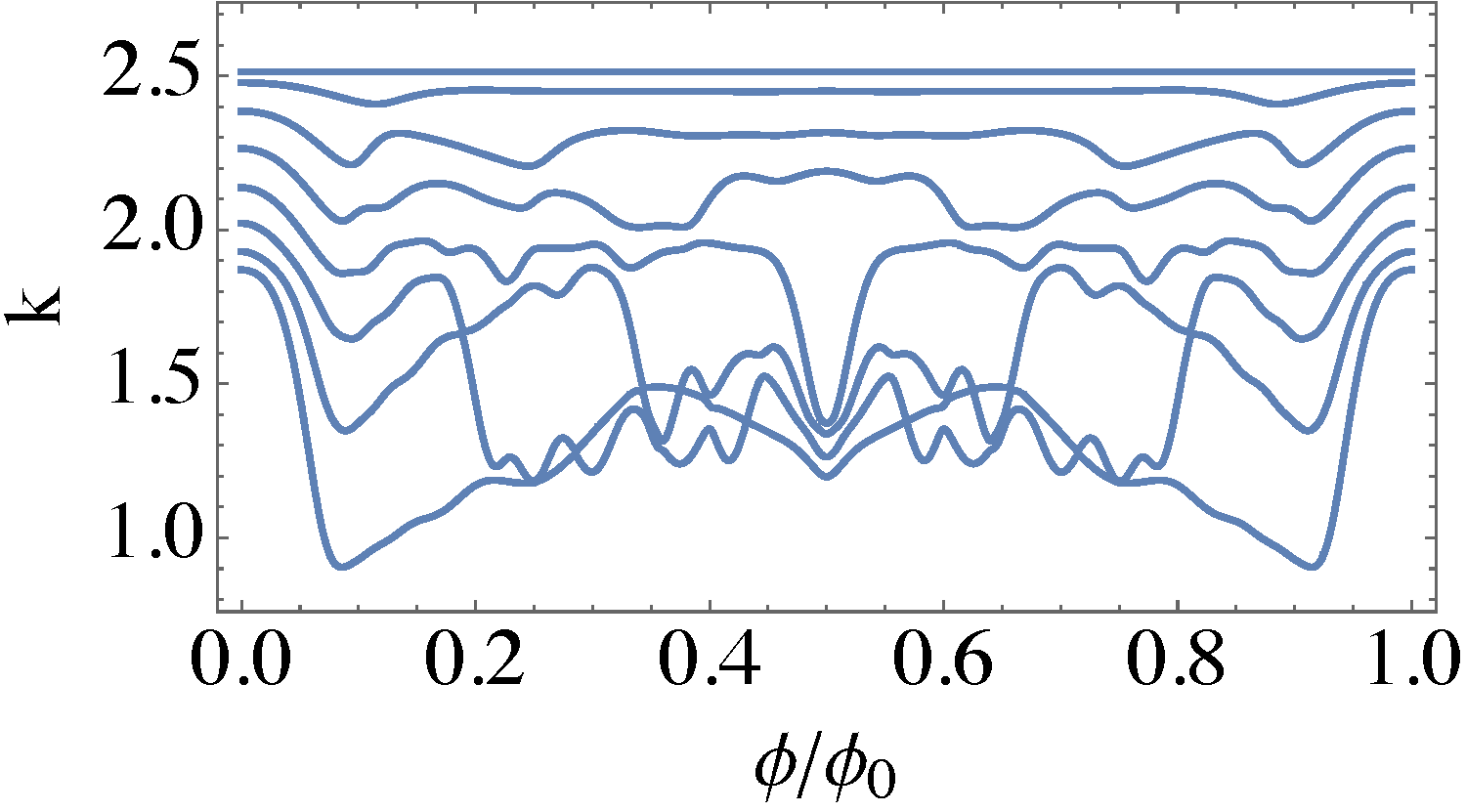}
{(d)~$M=13$}
\includegraphics[width=0.99\linewidth]{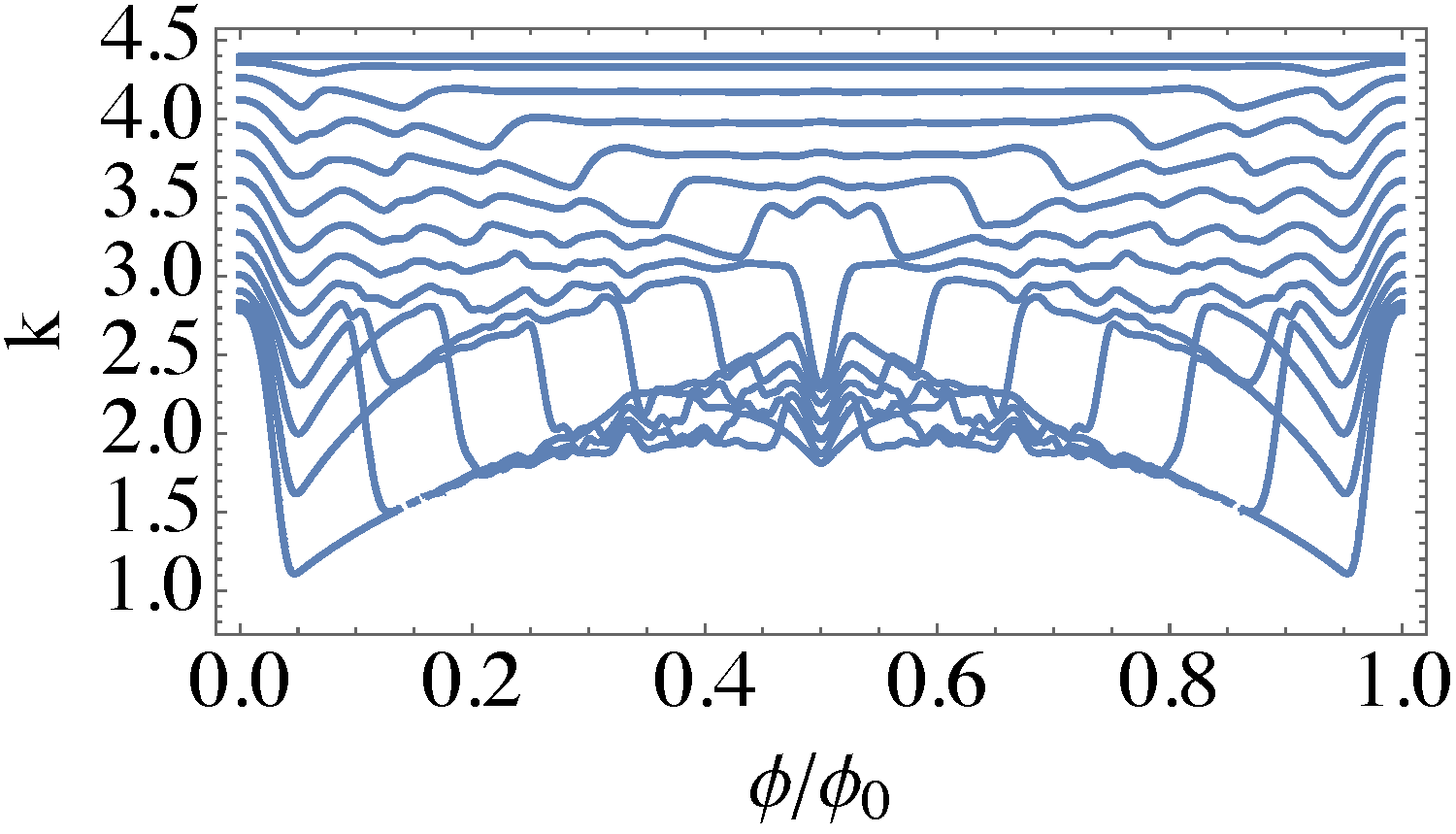}
\end{minipage}
\caption{
\label{fig:butterflies}
Quasimomentum spectra for systems with an increasing number of scatterers, whose heights are modulated according to eq.~\eqref{eq:cosine}. In all cases, $h_\textrm{min} = 0.1$ and $h_\textrm{max} = 2.1$. }
\end{figure}

We will first study the case of all positive scatterers, and fix the heights of the scatterers to vary between $h_\textrm{min} = 0.1$ and $h_\textrm{min} = 1.5$. 
We then find the quasimomentum spectrum for $\phi\in[0, \phi_0]$ and show it in \fref{fig:manyflies}(a) for $M=17$ scatterers.  
One can see that the spectrum is symmetric around $k=0$ and splits into bands, whose width depends on the minimum scatterer strength $h_\textrm{min}$.
Each of these bands has a shape that resembles a Hofstadter butterfly, but this shape becomes less pronounced in higher bands due to the finite height of the scatterers.
The gradual emergence of the butterfly-like shape with increasing the number of scatterers can be seen in \fref{fig:butterflies}.
One can also see that in each band the state with the largest absolute value of quasimomentum is fully flat and corresponds to a delocalized state with $k_\textrm{flat}^l = {\pi (M+1)l/L}$, where $l=\pm1,\pm2,...$ is the band index. This state has nodes exactly where the scatterers are located and its energy is therefore not affected by them.

Let us finally study the case which also includes negative values for the scatterers' strengths.
We limit ourselves to weak negative scatterers to avoid the presence of bound states.

The quasimomentum spectrum for $M=17$ scatterers with minimum and maximum strengths $h_\textrm{min} = -0.5$ and $h_\textrm{max} = 0.5$ is shown in Fig.~\ref{fig:manyflies}(b).
The two spectra in Fig.~\ref{fig:manyflies}(a) and (b) are very similar, especially for $k$ values close to the band-edge, which confirms again that the properties of the systems are mostly determined by the position distribution of the scatterers rather than their changes in strength. 
While the butterfly structure in this weakly-scattering, finite-sized system has not yet fully developed, one can see a prominent {\it cocoon}-shaped feature appearing around $k=0$ in the case where scatterers have negative as well as positive strengths.

\section{Conclusions}

We have used the coordinate Bethe ansatz to derive an analytical solution of the finite Kronig--Penney model with delta scatterers of arbitrary heights positioned at arbitrary points within a box.
The concise form of these solutions allows to treat many problems that were only accessible numerically until now in an exact way and is likely to give insight into many problems in solid state physics, such as impurities, finite systems, or disordered systems.

\rev{We showed that the bands of the KP model with uniform equidistant scatterers become topologically nontrivial upon applying a shift in the potential which represents a second virtual dimension. 
As a consequence of nontrivial topology, we observe chiral edge modes collapsing the gap in the analytical solution of our finite KP model.
Introducing random distortion in the heights of the scatterers, the gap remains open and the topology of the bands prevails.}

We have also demonstrated the appearance of a Hofstadter butterfly-like quasimomentum spectrum with modulated scatterer heights, as well as the presence of a cocoon-shaped feature in the spectrum in the case when the scatterers can be both positive and negative.
The solution we present can be readily applied to studies of localization in various distributions of the barrier heights and positions, in solid state and in optical lattices.

\section*{Acknowledgments}
This work was supported by the Okinawa Institute of Science and Technology Graduate University.


\end{document}